\documentclass[11pt]{article}
\usepackage{amsmath,amssymb,graphicx}
\newcommand{\ssy}[5]{#1  {\it #2}  {\bf #3} (#4) #5\rlap{.}}
\newcommand{\karti}[5][]{\begin{figure}[#2]\begin{center}%
\includegraphics[height=#3]{#4.ps}
\end{center}#1\caption{#5}\end{figure}}
\newcommand{\rmd}{\mathrm{d}}
\renewcommand{\leq}{\leqslant}

\DeclareMathOperator{\Ima}{Im}

\newcommand{\rea}{\text{I\!R}}
\title{What does the Letelier-Gal'tsov metric describe?}
\author{S. Krasnikov\\
\emph{The Central Astronomical Observatory at Pulkovo}}
\date{}
\begin{document}
\maketitle
\begin{abstract}
Recently the structure of the Letelier-Gal'tsov spacetime has become a
matter of some controversy. I show that the metric proposed in (Letelier
and Gal'tsov 1993 \emph{Class.\ Quantum Grav.}\ \textbf{10} L101) is
defined only on a part of the whole manifold. In the case where it can be
defined on the remainder by continuity, the resulting spacetime
corresponds to a system of parallel   cosmic strings at rest w.r.t. each
other.\\[\medskipamount] PACS: 04.20.Gz, 11.27.+d
\end{abstract}

\noindent Spacetimes with conical singularities, which presumably describe cosmic
strings, often exhibit a rich and non-trivial structure even when they
are flat. In studying such spacetimes it would be helpful to have a way
of constructing them in a uniform analytical way without resorting to
cut-and-glue surgery. One such a way was proposed some time ago by
Letelier and Gal'tsov. Consider a manifold $M=\rea^2\times P$, where $P$
is a plane, coordinatized by $x$ and $y$, from which $N$ points are cut
out. Endow $M$ with the metric
\begin{subequations}\label{eq:L-Gmetr}
\begin{equation}\label{eq:metrpr}
g:\qquad
\rmd s^2=\rmd t^2 - \rmd z^2 - \rmd Z\,\rmd \overline{Z},
\end{equation}
where
\begin{equation}\label{eq:defZ}
Z(\zeta,t) \equiv \int_{\zeta_0}^{\zeta} \prod_{i=1}^N
(\xi-\alpha_i(t,z))^{\mu_i}\rmd\xi,
\qquad \zeta\equiv x+ iy
\end{equation}
\end{subequations}
If $N=1$ and $\alpha_1=const$ the metric reduces to
\[
\rmd s^2=\rmd t^2 - \rmd z^2 - |\zeta-\alpha_1|^{2\mu_1}(\rmd x^2
 +\rmd y^2),
\]
which in the case $\mu_1>-1$ is the metric of a static   cosmic string
parallel to the $z$-axis. So, Letelier and Gal'tsov assumed \cite{letgal} that
in the general case the spacetime $(M,g)$ describes ``a system of crossed
straight infinite cosmic strings moving with arbitrary constant relative
velocities"
\cite{clegalet}. On the other hand, Anderson
\cite{and} calculated what he interprets as the distance between two strings
and found that it is constant. This led him to the conclusion that
\eqref{eq:L-Gmetr} ``is just the static parallel-string metric \dots written in
an obscure coordinate system''. Recently, however, his calculations have been
disputed in \cite{clegalet} where the opposite result was obtained. The goal
of the present note is to resolve this controversy. I show that in the general
case the metric
\eqref{eq:L-Gmetr} is defined only in \emph{a part} of $M$  and cannot be
extended to the remainder unless the resulting spacetime corresponds to
the set of  parallel strings at rest.

I shall consider the case $N=2$ (generalization to larger $N$ is trivial) and
for simplicity use the following notation
\[
 a\equiv\alpha_2(t_0),\quad
v\equiv\dot\alpha_2(t_0).
\]
I set
\[
\alpha_1(t_0)=\dot\alpha_1(t_0)=0,\qquad
\Ima a=\Ima v=0
\]
(this always can be achieved by an appropriate coordinate transformation and
thus involves no loss of generality). So, at $t=t_0$
\begin{equation}\label{eq:ourcaseZ}
Z(\zeta) =
\int_{\zeta_0}^{\zeta} \xi^{\mu_1}
(\xi-a)^{\mu_2}\rmd\xi,\quad Z,_t(\zeta)=-\mu_2v
\int_{\zeta_0}^{\zeta} \xi^{\mu_1} (\xi-a)^{\mu_2-1}\rmd\xi.
\end{equation}
Finally, I require that
\begin{equation}\label{eq:usnamu}
 \mu_1,\mu_2> -1,
\end{equation}
because if $\mu_i\leq -1$, the corresponding singularity is infinitely far and
does not represent a string.

Let us begin with the observation that
\[(\xi e^{2\pi i}-\alpha)^{\mu}\neq (\xi-\alpha)^{\mu}
,\qquad\text{when}\quad \mu\notin \mathsf Z\!\!\mathsf Z
\]
and therefore the function $Z$ and, correspondingly, the metric
\eqref{eq:metrpr} are generally
 not defined on the whole $M$. Let us choose
the domain $D_Z$ of $Z(\zeta)$ to be\footnote{This choice is made for the sake
of simplicity. Instead of $\rea_+$ one could take, say, a pair of curves
starting from 0 and $\alpha$, respectively. The result would not change.}
$P-\rea_+$. Then we can adopt the conventions that
\[
\zeta = |\zeta |e^{i\phi},\quad \zeta-a= |\zeta -a|e^{i\psi},\qquad
\forall\,\zeta\in D_Z,
\]
where $\phi$ and $\psi$ are the angles shown in figure~\ref{fig}, and
that the integrals in \eqref{eq:ourcaseZ} are taken along curves $\Gamma$
that do not intersect $\rea_+$. Substituting
\eqref{eq:ourcaseZ} in
\eqref{eq:metrpr} we find for any $\zeta\in D_Z$
\begin{equation*}
 g_{ty}(\zeta)= -\Ima\{\overline{Z},_\zeta Z,_t\}=
-\Ima\Big\{\overline{\zeta^{\mu_1} (\zeta-a)^{\mu_2}} Z,_t\Big\}.
\end{equation*}%
In particular, picking  $\zeta_*\in (0,a)$ we have
\begin{equation}\label{eq:gtx}
 g_{ty}(\zeta_*\pm i0)=
 -\zeta_*^{\mu_1}(a-\zeta_*)^{\mu_2}\Ima\Big\{e^{-i\pi[\mu_2+\mu_1(1\mp 1)]}
 Z,_t(\zeta_*\pm i0)\Big\}.
\end{equation}
Let us choose $\Gamma$ to be those of figure~\ref{fig}:
\[
Z,_t(\zeta_*\pm i0)=-\mu_2v
\int_{\Gamma^\pm} \xi^{\mu_1} (\xi-a)^{\mu_2-1}\rmd\xi.
\]
\karti{tb}{9.8 em}{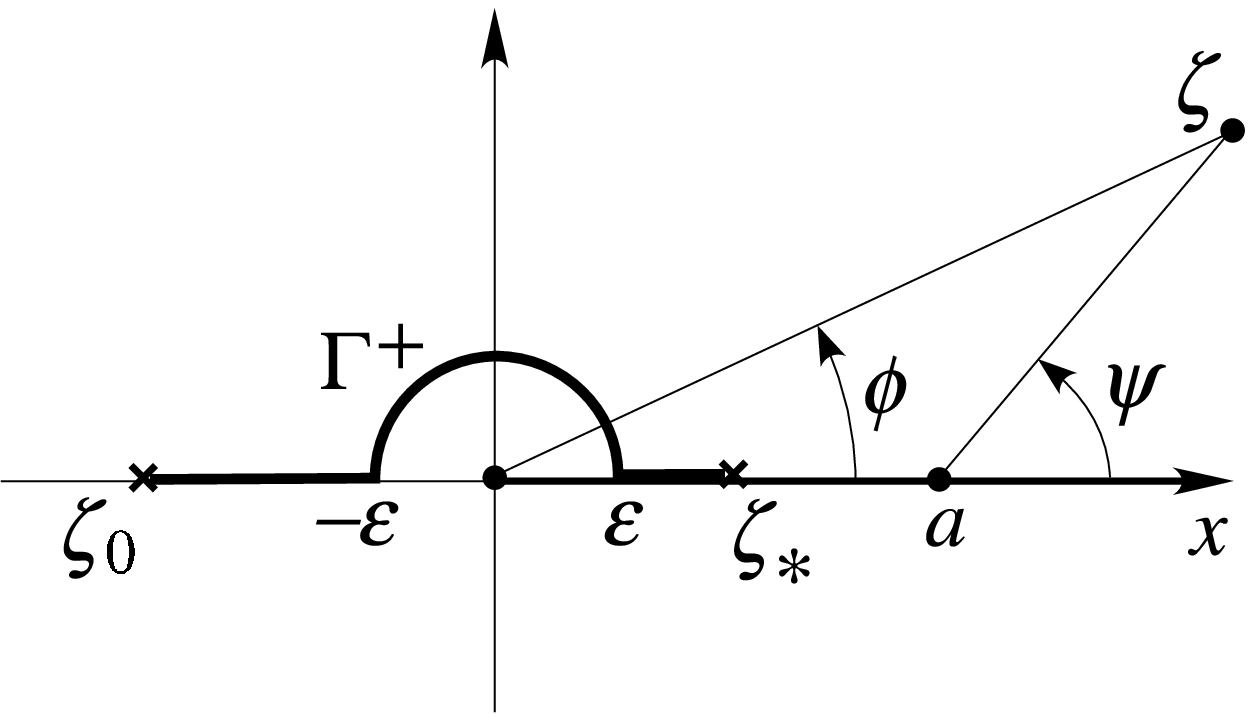}{\label{fig}  $\Gamma_+$ goes along the real axis from
$\zeta_0$, then along the upper half-circle of the radius $\varepsilon$, and
then along the upper bank of the cut.  $\Gamma_-$ is obtained from $\Gamma_+$
by
reflection w.r.t. the real axis.}%
The value of the integral does not depend on $\varepsilon$, while at
$\varepsilon\to 0$ the contribution of the half-circle vanishes by
\eqref{eq:usnamu}. Hence
\begin{multline*}
Z,_t(\zeta_*\pm i0)= e^{i\pi(\mu_1+\mu_2)}\mu_2v\int^{|\zeta_0|}_0 r^{\mu_1}
(r+a)^{\mu_2-1}\rmd r
\\
+ e^{i\pi[\mu_2+\mu_1(1\mp 1)]}\mu_2v\int^{\zeta_*}_0 r^{\mu_1}
(a-r)^{\mu_2-1}\rmd r,
\end{multline*}
which being substituted in \eqref{eq:gtx} gives
\begin{multline*}
 g_{ty}(\zeta_*\pm i0)
=-\mu_2v\zeta_*^{\mu_1}(a-\zeta_*)^{\mu_2}
 \Ima\Big\{e^{\pm i\pi\mu_1}
 \int^{|\zeta_0|}_0 r^{\mu_1}(r+a)^{\mu_2-1}\rmd r
 \\
 +\int^{\zeta_*}_0 r^{\mu_1}
(a-r)^{\mu_2-1}\rmd r\Big\}
\end{multline*}
and as a result
\[
g_{ty}(\zeta_*+i0)-g_{ty}(\zeta_*- i0)= -2
\mu_2 v\zeta_*^{\mu_1}(a-\zeta_*)^{\mu_2}
\sin(\pi\mu_1)
 \int^{|\zeta_0|}_0 r^{\mu_1}(r+a)^{\mu_2-1}\rmd r.
\]
Since the metric must be continuous, for a negative $\mu_1$ it follows $v=0$,
 i.~e.\ the strings are at rest with respect to each other. Repeating exactly the
same reasoning with $t$ changed to $z$ one finds that
$\partial_z\alpha_2(t_0)=0$ and so, the strings are also parallel.
\paragraph{Remark.} If one allows $\mu_1$ to be positive, an exception appears:
the metric may be smooth for $v\neq 0$ if $\mu_1= n$. The singularities of
this type are interesting and useful \cite{HS}, but  can hardly be called
`strings' (in particular such singularities cannot be `smoothed out' without
violation of the weak energy condition).

\end{document}